\documentclass[apj]{emulateapj}
\begin{document}
\newcommand{\msun}{\mbox{M$_{\odot}$}}
\newcommand{\rsun}{\mbox{R$_{\odot}$}}
\newcommand{\zsun}{\mbox{Z$_{\odot}$}}
\newcommand{\lsun}{\mbox{L$_{\odot}$}} 

\title{The Distance to the Massive Eclipsing Binary LMC-SC1-105 \\ in the
  Large Magellanic Cloud}

\author{Alceste Z. Bonanos\altaffilmark{1}, Norberto
  Castro\altaffilmark{1}, Lucas M. Macri\altaffilmark{2}, Rolf-Peter
  Kudritzki\altaffilmark{3,4}}

\altaffiltext{1}{Institute of Astronomy \& Astrophysics, National
  Observatory of Athens, I. Metaxa \& Vas. Pavlou St., P. Penteli, 15236
  Athens, Greece; bonanos@astro.noa.gr, norberto@noa.gr}
\altaffiltext{2}{George P. and Cynthia Woods Mitchell Institute for
  Fundamental Physics and Astronomy, Department of Physics \& Astronomy,
  Texas A\&M University, 4242 TAMU, College Station, TX 77843-4242, USA;
  lmacri@physics.tamu.edu} \altaffiltext{3}{Institute for Astronomy,
  University of Hawaii, 2680 Woodlawn Dr., Honolulu, HI 96822, USA;
  kud@ifa.hawaii.edu}\altaffiltext{4}{Max-Planck-Institute for
  Astrophysics, Karl-Schwarzschild-Str. 1, D-85741 Garching, Germany.}

\begin{abstract}

This Letter presents the first distance measurement to the massive,
semi-detached, eclipsing binary LMC-SC1-105, located in the LH~81
association of the Large Magellanic Cloud (LMC). Previously determined
parameters of the system are combined with new near-infrared photometry
and a new temperature analysis to constrain the reddening toward the
system, and determine a distance of $50.6\pm1.6$ kpc (corresponding to a
distance modulus of $18.52\pm0.07$ mag), in agreement with previous
eclipsing binary measurements. This is the sixth distance measurement to
an eclipsing binary in the LMC, although the first to an O-type
system. We thus demonstrate the suitability of O-type eclipsing binaries
(EBs) as distance indicators. We suggest using bright, early-type EBs to
measure distances along different sight lines, as an independent way to
map the depth of the LMC and resolve the controversy about its
three-dimensional structure.

\end{abstract}


\section{Introduction}
\label{section:intro}

%

As one of the nearest galaxies to the Milky Way, the Large Magellanic
Cloud (LMC) has naturally been an attractive first rung for the
Extragalactic Distance Scale. The {\it HST} Key Project
\citep{Freedman01} adopted a distance modulus $\mu=18.50\pm0.10$ mag
(corresponding to a distance of 50.1$\pm2.4$ kpc) to the LMC, which has
since become the consensus in the community. \citet{Schaefer08} pointed
out that overestimation of error bars and band-wagon effects are present
in the literature, with pre-2001 LMC distance measurements yielding
values between 18.1 and 18.8 mag \citep[see][]{Benedict02}, and
post-2001 values clustering around the Key Project value. Given that
different systematic errors accompany each method, a careful comparison
of the distances resulting from different methods is necessary to
characterize them. Furthermore, there is increasing evidence for
substantial and complex vertical structure in the disk of the LMC
\citep[see review by][]{vanderMarel06} from studies of red clump stars
\citep{Olsen02,Subramanian10}, Cepheid variables \citep{Nikolaev04} and
RR Lyrae stars \citep{Pejcha09}, which demands further exploration.

The only direct, geometrical method available for measuring distances to
stars in the LMC is with eclipsing binaries (EBs). In particular, the
light curve provides the fractional radii of the components, the radial
velocity semi-amplitudes determine the masses and size of the orbit,
which together with the effective temperature determination (e.g.\ by
comparison with synthetic spectra), yield luminosities and therefore
distances \citep[see reviews by][]{Andersen91,Torres10}. The EB distance
method has so far been applied to four early-B type systems
\citep{Guinan98, Ribas02, Fitzpatrick02, Fitzpatrick03} and one G-type
giant system \citep{Pietrzynski09} in the LMC, with individual
uncertainties ranging from 1.2 to 2.2 kpc. Four of these systems are
located within the bar of the LMC and their individual distances are
consistent with the quoted uncertainties, yielding an error-weighted
mean value of $49.4\pm1.1$~kpc. A fifth system, located several degrees
away in the north-east quadrant of the disk of the LMC, gives a
$3\sigma$ shorter distance of $43.2\pm1.8$~kpc.

\begin{figure}[ht]  
\includegraphics[width=8.5cm]{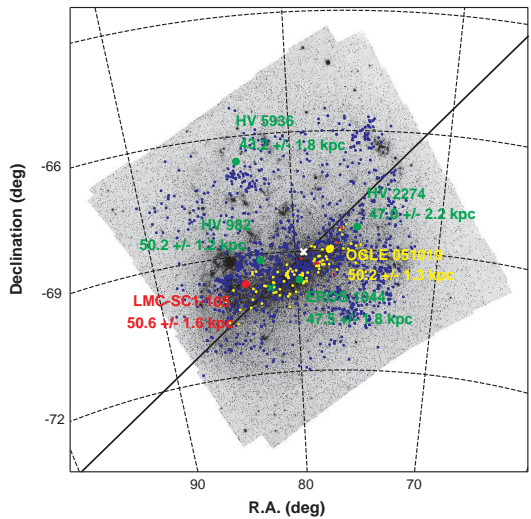}
\caption{Spatial distribution of known EBs from OGLE II and MACHO (blue
  circles) on the {\it Spitzer} 3.6$\mu$m image of the LMC. EBs with
  measured distances are labeled.  Yellow circles mark the most suitable
  detached EBs for distance determination \citep{Michalska05}; red
  circles mark the OGLE II binaries we plan to measure distances to
  next. The \ion{H}{1} kinematic center (white ``x'') from \citet{Kim98}
  and the dynamical center or center of the bar (green ``x'') from
  \citet{vanderMarel02} are labeled; the solid line corresponds to the
  line of nodes \citep{vanderMarel02}. Coordinates are given for J2000.}
\label{map}
\end{figure}

Figure~\ref{map} shows the spatial distribution of all known EBs from
the OGLE II \citep{Wyrzykowski03} and MACHO \citep{Derekas07,
  Faccioli07} microlensing surveys of the LMC, and the systems with
measured distances, overlaid onto the {\it Spitzer} SAGE image in the
IRAC 3.6 $\mu$m band \citep{Meixner06}. A magnitude cut ($V<17$ mag) and
period cut ($>1.5$ days) were both applied to the EB catalogs to reject
foreground systems and faint systems whose immediate follow up is
unrealistic or impossible. The detached EBs selected by
\citet{Michalska05} among the OGLE II systems as being most suitable for
distance determination are also shown. Both the \ion{H}{1} kinematic
center \citep{Kim98} and the dynamical center \citep[or center of the
  bar;][]{vanderMarel02} are overplotted, as is the line of nodes
\citep[$\Theta=129.\!^{\circ}9\pm6.\!^{\circ}0\deg$;][]{vanderMarel02}.

Motivated by the evidence for vertical structure in the LMC and the one
discrepant EB distance, we proceed to compute the distance to
LMC-SC1-105\footnote{Or OGLE J053448.26-694236.4 = MACHO 81.8881.21 =
  LH~81-72.}. LMC-SC1-105 is a massive, semi-detached, short period
($P=4.25$ days) O-type system, with component masses of $\rm
M_{1}=30.9\pm1.0\;\msun$, $\rm M_{2}=13.0\pm0.7\;\msun$, and radii of
$\rm R_{1}=15.1\pm0.2\;\rsun$, $\rm R_{2}=11.9\pm0.2\;\rsun$
\citep[determined by][]{Bonanos09}. The very accurate measurement of the
radii ($<2\%$) renders the system suitable for a distance determination,
given that EB distances are independent of the usual distance ladder and
therefore important checks for other methods. However, accurate radii
are not sufficient for an accurate distance. Accurate fluxes
(i.e.\ effective temperatures) and extinction estimates are also needed,
therefore this Letter sets out to determine these quantities and obtain
the distance. Specifically, Section 2 presents new near-infrared
photometry of LMC-SC1-105, Section 3 an analysis of the spectra with
state-of-the-art model atmospheres, Section 4 the distance
determination, and finally, Section 5 a discussion of our results.

\section{Near-Infrared Data}

\begin{figure}[ht]
\includegraphics[width=8.5cm]{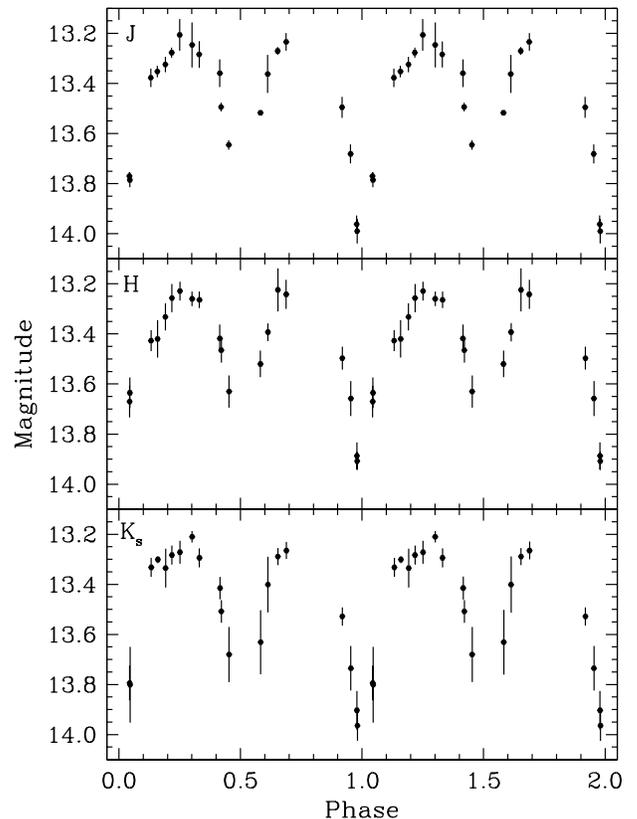}
\caption{Phased CPAPIR $JHK_s-$band light curves of LMC-SC1-105.}
\label{lcs}
\end{figure}

This study makes use of $JHK_s$ observations of LMC-SC1-105 obtained
with the CPAPIR camera \citep{Artigau04} at the CTIO 1.5-m, as part of a
synoptic survey of Cepheid variables in the LMC (L. M. Macri et
al. 2011, in prep.). The EB was observed at 20 different epochs on 11
nights between 2006 November 5 and 2007 December 2. Time-series PSF
photometry was carried out using DAOPHOT and ALLFRAME
\citep{Stetson87,Stetson94}. Photometric zeropoints were determined
using $\sim 2,500$ stars from the 2MASS Point Source Catalog, located
within $15\arcmin$ of the system and with $10.5<K_s<13.5$~mag, while
color terms were derived using nearly $5\times 10^5$ 2MASS stars across
the entire bar of the LMC. Figure~\ref{lcs} shows the calibrated, phased
light curves of LMC-SC1-105. We calculated error-weighted out-of-eclipse
mean magnitudes of $J=13.22\pm0.04$, $H=13.27\pm0.04$ and
$K_s=13.26\pm0.04$~mag.

\section{Effective Temperature Analysis}

An accurate distance measurement to LMC-SC-105 requires an accurate flux
determination for its binary components. We proceed to refine the
effective temperatures estimated by \citet{Bonanos09}\footnote{T$_{\rm
    eff1}=35\pm2.5$kK, T$_{\rm eff2}=32.5\pm2.5$kK, for $\log(g)=3.50$
  (fixed), from best fit TLUSTY models \citep{Lanz03}.}  with the
state-of-the-art, NLTE stellar atmosphere code FASTWIND
\citep{Santolaya97,Puls05}, which includes the effects of stellar winds
and spherical atmospheric extension.

The analysis involves a direct comparison between the helium lines,
which are the main temperature diagnostics at these spectral types, plus
H$\alpha$, to constrain the stellar wind, with a complete FASTWIND model
grid designed to study O-type stars at the metallicity of the LMC. The
grid was developed within the FLAMES-II collaboration \citep{Evans10}
and constructed at the Instituto de Astrof\'{i}sica de
Canarias. Specifically, we derived the set of models that provide the
lowest $\chi^2$, using H$\alpha$ and the 10 strongest \ion{He}{1} and
\ion{He}{2} lines available\footnote{\ion{He}{1} $\lambda\lambda$4026,
  4143, 4471, 4713, 4922, 5015, 5875 and \ion{He}{2}
  $\lambda\lambda$4200, 4541, 5411.}. The synthetic models were
downgraded to the instrumental resolution of the observed spectra and
the projected rotational velocities $v \sin i$ were refined to 160
km~s$^{-1}$ and 120 km~s$^{-1}$, for the primary and secondary
component, respectively. We fixed the surface gravities to the values
determined by \citet{Bonanos09}: $\log(g_1)=3.57\pm0.02$ and
$\log(g_2)=3.40\pm0.03$\footnote{Note, the $\log(g)$ error bars given in
  Table~5 of \citet{Bonanos09} incorrectly correspond to the errors in
  $g$.}. In practice, we rounded the values to the first decimal point,
to match the 0.1 dex step size of the grid. The $\chi^2$ method provides
the stellar parameters and their corresponding errors.

\begin{figure}[ht]  
\includegraphics[angle=90,width=8.5cm]{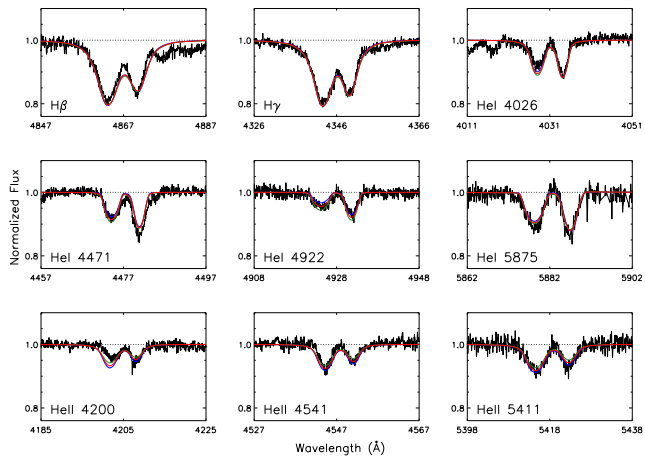}
\caption{Best fit FASTWIND model (red) of LMC-SC1-105, at the first
  quadrature. The blue (green) lines correspond to models with the best
  fit T$\rm_{eff}$ plus (minus) the 1$\sigma$ error. The set of lines
  with smaller Doppler shifts corresponds to the primary.}
\label{oct23}
\end{figure}

\begin{figure}[ht]  
\includegraphics[angle=90,width=8.5cm]{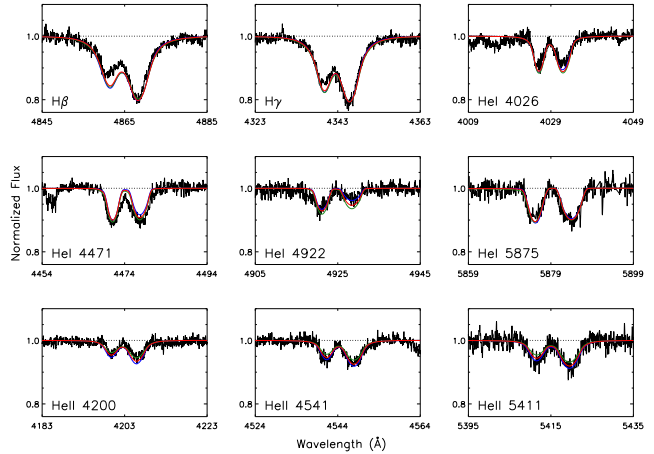}
\caption{Same as Figure~\ref{oct23}, but for the second quadrature. The
  set of lines with larger Doppler shifts corresponds to the primary.}
\label{oct25}
\end{figure}

The technique was applied to the two highest S/N spectra of LMC-SC1-105
\citep[see][]{Bonanos09}, obtained at phases 0.27 and 0.75, i.e.\ at the
first and second quadratures. Both phases yielded the same temperature
for each component, within the errors. Specifically, at first
quadrature, we found best fit values of T$_{\rm eff1}=36100\pm1000$~K,
T$_{\rm eff2}=33200\pm800$~K, while at the second quadrature T$_{\rm
  eff1}=35700\pm1100$~K, T$_{\rm
  eff2}=33100\pm900$~K. Figures~\ref{oct23} and \ref{oct25} show the
best fit FASTWIND models, plus the effects of the temperature errors in
the profiles. The synthetic models, which only include transitions of
\ion{H}{1}, \ion{He}{1} and \ion{He}{2}, provide a good match to the
observed spectra. Despite not including the Balmer lines in the analysis
(except H$\alpha$), the wings of these lines are in good agreement with
the models, confirming the accuracy of the $\log(g)$ determination from
the EB analysis.

\citet{Bonanos09} reported changes of the spectral types with phase due
to the Struve-Sahade effect \citep{Stickland97}, the largest being from
O7V to O8V for the primary, which would have an impact on the
temperature of $\sim2000$~K \citep{Martins05}. Our analysis, however,
does not yield any remarkable differences in temperature between the two
quadratures. The reason for this is that the classification criteria
\citep{Walborn90} hinge on the lines \ion{He}{2}~$\lambda4541$,
\ion{He}{1}~$\lambda4471$, \ion{He}{2}~$\lambda4200$, and
\ion{He}{1}+II~$\lambda4026$, while the FASTWIND analysis averaged over
10 \ion{He}{1} and \ion{He}{2} lines in the spectrum. The imperfect fits
of \ion{He}{2}~$\lambda4200$ and \ion{He}{1}~$\lambda4471$ by the models
(see Figure~\ref{oct23}), are consistent with a spectral type change.

At phase 0.75, the secondary star shows important deviations in the
cores of H$\beta$ and H$\gamma$ from the model (see Figure~\ref{oct25}),
which might be due to excess emission arising from the slow mass
transfer or the distorted line profiles of Roche lobe-filling stars
\citep[see][]{Bitner06}. Nonetheless, the rest of the \ion{He}{1} and
\ion{He}{2} lines are well modeled within the errors. Some of the
\ion{He}{2} lines (e.g.\ \ion{He}{2}~$\lambda$4541) might indicate a
slightly higher temperature, however these differences lie within the
errors.

\section{Distance}

The flux $f_{\lambda}$ measured at Earth at a certain wavelength
$\lambda$ from a binary at distance $d$ is given by

\begin{equation}
\label{disteq}
f_{\lambda}=\frac{1}{d^2} \left(R_{1}^2 \; F_{1,\lambda}+R_{2}^2 \;
F_{2,\lambda} \right) \times 10^{-0.4 \; A\left(\lambda \right)},
\end{equation} 

\noindent where $R_{1}$ and $R_{2}$ are the radii of the two stars and
$F_{1,\lambda}$ and $F_{2,\lambda}$ the surface fluxes. The total
extinction $A\left(\lambda \right)$ is a function of the reddening
$E(B-V)$, the normalized extinction curve $k(\lambda-V) \equiv
E(\lambda-V)/E(B-V)$ and the ratio of total to selective extinction in
the $V$ band, $R_V \equiv A(V)/E(B-V)$:

\begin{equation}
A\left(\lambda \right) =
E\left(B-V\right)\left[k\left(\lambda-V\right)+R_V\right].
\end{equation} 

Having measured the temperatures of the stars from the spectra, we
computed fluxes and fit to the observed magnitudes, using Equation
\ref{disteq} and the best-fit FASTWIND model atmospheres for each
quadrature determined above. Note that we used the mean
radii\footnote{$\rm (r_{pole}+r_{side}+r_{back})/3$} of the stars
instead of their volume radii as better approximations to compute their
projected surface areas.

Following the procedure outlined in \citet{Bonanos06} for the detached
EB in M33, we calculated synthetic photometry of the composite spectrum
over the appropriate Johnson-Cousins optical filter functions as defined
by \citet{Bessell90} and calibrated by \citet{Landolt92}, and the 2MASS
filter set. Monochromatic fluxes were measured at the isophotal
wavelengths \citep[see][]{Tokunaga05}, which best represent the flux in
a passband. We used zeropoints from \citet[][Appendix A]{Bessell98} and
\citet{Cohen03} to convert the fluxes to magnitudes. We reddened the
model spectrum using the reddening law parameterization of
\citet{Cardelli89}, as prescribed in \citet{Schlegel98}, and
simultaneously fit the optical\footnote{$B_{\rm max}=12.81\pm0.01$~mag,
  $V_{\rm max}=12.97\pm0.01$~mag, $I_{\rm max}=13.04\pm0.01$~mag
  \citep{Wyrzykowski03}.} and near-infrared $BVIJHK_s$
photometry. Specifically, we computed the intrinsic $(B-V)_0=-0.27$ mag
from the model atmospheres at the isophotal wavelengths, thus yielding
$E(B-V)=0.11\pm0.01$ mag.

\begin{figure}[ht]  
\includegraphics[angle=90,width=8.5cm]{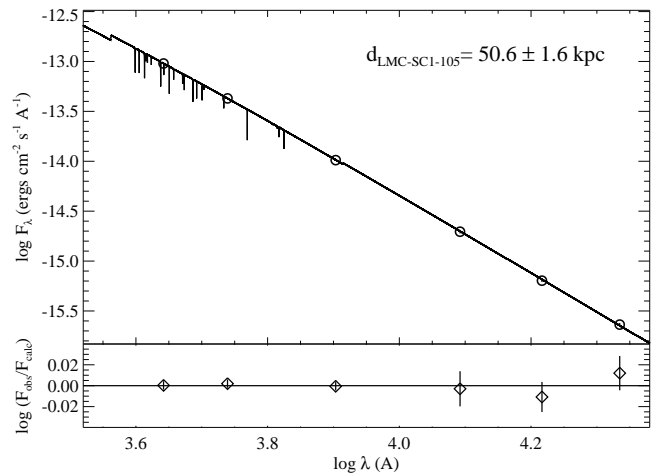}
\caption{Upper panel: fit of the reddened EB model spectrum (for phase
  0.27) to the $BVIJHK_s$ photometry. Lower panel: residuals of the SED
  fit, in terms of the flux ratio. Error bars correspond to the
  photometric error for each band in flux units. The best fit values of
  $E(B-V)=0.11\pm0.01$ mag and $R_{V}=5.8\pm0.4$ yield a distance
  modulus to the EB and thus the LMC bar of $50.6\pm1.6$ kpc
  ($\mu=18.52\pm0.07$ mag).}
\label{sed}
\end{figure}

The value of $R_V$ was determined as the value that minimized the error
in the SED fit over the six photometric bands. For phase 0.27, we found
$R_V=5.8\pm0.4$ and for phase 0.75, $R_V=5.7\pm0.4$. The resulting
distance to LMC-SC1-105 and thus the LMC bar is $50.6\pm1.6$ kpc
($\mu=18.52\pm0.07$ mag) for the first quadrature and $50.4\pm1.6$ kpc
($\mu=18.51\pm0.07$ mag) for the second quadrature. The distances are
identical within errors. Given the better fit of the FASTWIND models to
the spectra at first quadrature, we adopt the distance derived for first
quadrature. The fit of the reddened model spectrum to the photometry and
the residuals of the fit are shown in the upper and lower panels of
Figure~\ref{sed}, respectively. The error in the distances was computed
by a bootstrap resampling procedure. We repeated the spectral energy
distribution (SED) fitting procedure 1000 times for each quadrature, by
randomly selecting (using Gaussian sampling) all the parameters within
their errors. We adopt the $\sigma$ of the resulting Gaussian
distribution as the uncertainty in the distance.

We tested the robustness of our reddening and distance results, by first
fitting the $BVI$ photometry alone, which yielded an identical value for
the distance ($50.8\pm1.6$ kpc or $\mu=18.53\pm0.07$ mag, with
$R_V=5.7\pm0.4$), thus demonstrating the consistency of the
near-infrared with the optical photometry. Next, if we fix $R_V=3.1$,
the best fit value for $E(B-V)=0.18$ mag, resulting in a distance of
$51.9\pm1.6$ kpc ($\mu=18.58\pm0.07$ mag), i.e.\ in agreement with our
reported result, within errors. If instead we assume $R_V=3.1$ and fix
$E(B-V)=0.11\pm0.01$ mag (based on our photometry and the model
spectra)\footnote{Note, our $E(B-V)$ value is consistent with the range
  (0.13-0.23 mag) measured by \citet{Massey00} for 34 stars in LH~81.},
we would derive a much larger distance of 55.2 kpc ($\mu=18.71$ mag),
which yields a SED fit error of 0.05 mag (versus 0.01 mag) that is
inconsistent with the photometry. The validity and implications of the
high value of $R_V$ that we have measured are discussed in the following
Section.

The error quoted above for $R_V$ was estimated using the Bayesian code
CHORIZOS \citep{Maiz-Apellaniz04}. The available $BVIJHK_s$ photometry
was given as input, with Teff in the range 33000$-$36000~K and $\log(g)$
fixed to 3.50, from TLUSTY models. The code yielded best fit mean values
(for a single star) of T$_{\rm eff}=34500\pm1100$~K,
$R_{\lambda5495}=5.4\pm0.4$ and $E(\lambda4405-\lambda5495)=0.10\pm0.01$
mag, consistent with the values we derived.

\section{Discussion}

LMC-SC1-105 is located in the LH 81 association \citep{Massey00}, near
the center of the LMC bar. It contains two early O-type stars and three
Wolf-Rayet systems, one of which was recently found to be an EB
\citep{Szczygiel10}. Furthermore, this association resides in the
superbubble N 154 \citep{Henize56} = DEM 246 \citep{Davies76}. We have
determined a large value of $R_V=5.8\pm0.4$ toward LMC-SC1-105, however,
such high values are not uncommon. \citet{Cardelli89} find
$5<R_V\leq5.6$ for 6 out of the 29 OB stars in their sample, while
\citet{Fitzpatrick07} find $R_V>5$ for 12 out of the 328 stars in their
sample. Large values of $R_V$ simply imply larger dust grain sizes,
which are expected to occur in dense regions of the interstellar medium
due to accretion and coagulation of grains. We therefore conclude that
the environment in which LMC-SC1-105 resides has large dust grains.

In this Letter, we have determined the distance to LMC-SC1-105 and
consequently the LMC bar to be $50.6\pm1.6$ kpc ($\mu=18.52\pm0.07$
mag). The agreement we find with previous EB distances to systems in the
bar with different spectral types testifies to the robustness of the EB
method and its potential as a powerful, independent distance
indicator. Furthermore, it confirms that O-type (and semi-detached) EBs
are suitable for distance determination, i.e.\ that the fluxes predicted
by FASTWIND are indeed accurate. EB-based distance determinations to M31
\citep{Ribas05, Vilardell10} and M33 \citep{Bonanos06} can therefore
provide an independent absolute calibration of the Extragalactic
Distance Scale. Future distance determinations to EBs in the LMC
(e.g.\ those marked in Figure~\ref{map}), will additionally provide
$R_V$ values in different environments of the LMC. Finally, we suggest
using bright, early-type EBs to measure distances along different sight
lines to the LMC, as an independent way to map its depth and resolve the
controversy about its vertical structure.

\acknowledgments{We are very grateful to S. S\'{i}mon-D\'{i}az for
  making available part of his FASTWIND grid at $Z/Z_\odot=0.4$.
  L.M.M. thanks Shashi Kanbur and Chow-Choong Ngeow for allowing the use
  of the CPAPIR data in advance of publication. The CPAPIR survey of the
  LMC was made possible by faculty startup funds from the State
  University of New York at Oswego and Texas A\&M University. A.Z.B. and
  N.C.  acknowledge research and travel support from the European
  Commission Framework Program Seven under a Marie Curie International
  Reintegration Grant. R.P.K. acknowledges support by the
  Alexander-von-Humboldt-Foundation and from the National Science
  Foundation under grant AST-1008797. This research has made use of
  SAOImage DS9, developed by Smithsonian Astrophysical Observatory.}



\end{document}